\documentclass[a4paper,11pt]{article}

\usepackage{contribution}

\newcommand{\weblink}[2][]{%
    \ifthenelse{\equal{#1}{}}%
    {\textnormal{\url{#2}}}%
    {\textnormal{\href{#2}{#1}}}%
}

\newcommand{\acknowledgements}[1]{%
  \bigskip\bigskip
  \textsf{\textbf{\Large Acknowledgements}} \\[2ex]
  {#1}
  \bigskip
}

\def\beq{\begin{equation}}
\def\eeq#1{\label{#1}\end{equation}}
\def\eeqn{\end{equation}}

\def\beqa{\begin{eqnarray}}
\def\eeqa#1{\label{#1}\end{eqnarray}}
\def\eeqan{\end{eqnarray}}

\let\bar=\overbar

\def\etal{{\it et al.}}

\def\Dslash{\not{\hbox{\kern-4pt $D$}}}
\def\dslash{\not{\hbox{\kern-2pt $\del$}}}

\def\msb{{\bar{\ssstyle M \kern -1pt S}}}

\newcommand{\contribution}[7][]{%
  \clearpage
  \thispagestyle{plain}
  \ifthenelse{\equal{#1}{}}
  {\hypersetup{pdftitle={#2}}}
  {\hypersetup{pdftitle={#1}}}
  \hypersetup{pdfauthor={{#3} {#4}}}
  {\centering\normalfont\LARGE\bfseries\sffamily #2 \par\nobreak}
  \lhead{}
  \chead{%
    \textit{\footnotesize XIV International Conference on Hadron Spectroscopy
      (\weblink[\textit{hadron2011}]{http://www.hadron2011.de}), 13-17 June 2011, Munich, Germany}%
  }
  \rhead{}
  \bigskip
  \begin{center}
    {#3} {#4}\ifthenelse{\equal{#6}{}}{}{\footnote{\weblink[#6]{mailto:#6}}}
    \ifthenelse{\equal{#7}{}}{}{#7} \\
    \textit{#5}
  \end{center}
  \bigskip
}

\renewcommand{\abstract}[1]{%
  \begin{center}
    \begin{minipage}{0.85\textwidth}
      \begin{footnotesize}
        #1
      \end{footnotesize}
    \end{minipage}
  \end{center}
  \bigskip
}

\begin{document}

%
%
%
%
%
{  

\newcommand{\Journal}[4]     {{#1} {\bf {#2}}, {#3} ({#4})}
\newcommand{\EPJ}            {Eur.\ Phys.\ J.\ }
\newcommand{\JINST}          {JINST}
\newcommand{\JHEP}           {JHEP}
\newcommand{\PhysLett}       {Phys.\ Lett.\ }
\newcommand{\PhysRev}        {Phys.\ Rev.\ }
\newcommand{\PRL}            {Phys.\ Rev.\ Lett.\ }
\newcommand{\NuclPhys}       {Nucl.\ Phys.\ }
\newcommand{\CompPhysComm}   {Comp.\ Phys.\ Commun.\ }

\newcommand{\WWWAddr}[1]     {{\tt {#1}}}
\newcommand{\hepex}[1]       {{\tt hep-ex/{#1}}}
\newcommand{\arXiv}[2]       {{\tt arXiv:{#1}.{#2}}}   
\newcommand{\arXivOld}[2]    {{\tt arXiv:{#1}/{#2}}}   
\newcommand{\ATLASConf}[2]   {ATLAS-CONF-{#1}-{#2}}

\newcommand{\unitexp}[2]  {\mbox{{#1}$^{\mathrm{#2}}$}}
\newcommand{\scinot}[2]   {\mbox{{#1}$\times$10$^{\mathrm{#2}}$}}
\newcommand{\abs}[1]      {\mbox{$\mid {#1} \mid$}}
\newcommand{\order}[1]    {\mbox{$\mathcal{O}$(#1)}}
\newcommand{\BRat}[1]     {\mbox{$\mathcal{B}$(#1)}}

\newcommand{\Pythia}   {\mbox{\tt PYTHIA}}
\newcommand{\Herwig}   {\mbox{\tt HERWIG}}
\newcommand{\POWHEG}   {\mbox{\tt POWHEG}}
\newcommand{\MCatNLO}  {\mbox{\tt MC@NLO}}

\newcommand{\instLunit}{\mbox{\unitexp{cm}{-2}$\cdot$\unitexp{s}{-1}}}
\newcommand{\um}       {\mbox{$\mu$m}}
\newcommand{\ra}       {\mbox{$\rightarrow$}}

\newcommand{\pT}       {\mbox{$p_T$}}
\newcommand{\dR}       {\mbox{$\Delta R$}}
\newcommand{\sqrts}    {\mbox{$\sqrt{s}$}}

\newcommand{\qqb}[1]   {\mbox{${#1} \bar{#1}$}}
\newcommand{\mumu}     {\mbox{$\mu^+ \mu^-$}}
\newcommand{\pp}       {\mbox{$pp$}}
\newcommand{\Jpsi}     {\mbox{$J/\psi$}}
\newcommand{\Bs}       {\mbox{$B_s$}}

\contribution[Heavy Hadrons at the LHC]  
{Production and Spectroscopy of Heavy Hadrons at the LHC}  
{Harold}{Evans}  
{Physics Department \\
  Indiana University \\
  Bloomington, IN, 47405-7105\\ 
  USA}  
{hgevans@indiana.edu}  
{on behalf of the ALICE, ATLAS, CMS, and LHCb Collaborations}  

\abstract{%
  Measurements of heavy flavor production and decay have featured
  prominently in the early results from the four large LHC
  experiments: ALICE, ATLAS, CMS, and LHCb. These results provide
  tests of QCD models in a new energy region and point the way toward
  future measurements of $CP$ violation and searches for new
  physics. An overview of open heavy flavor studies is presented here,
  focusing 
  on how the new measurements extend our knowledge of this area of
  physics. 
  Heavy quarkonia states at the LHC are summarized in other
  proceedings of this conference.
  I also discuss briefly how heavy flavor measurements are
  likely to evolve as LHC luminosities increase.
}

\section{Introduction}

Heavy hadron production and spectroscopy are not primary focuses of any of
the four large LHC experiments. Nevertheless, each of the
collaborations have produced interesting and important results in this
area. In fact, more than 50 heavy flavor results were available from
ALICE, ATLAS, CMS, and LHCb at the time of this conference. Clearly, I
will not be able to cover them all in this brief
review. Rather, I'll attempt to concentrate on large themes in the
field and discuss how recent results from the LHC have contributed to
our understanding in these areas. 
Necessarily, I will be forced to gloss over many interesting details
of the measurements I mention.
More information on theses measurements can be found in the
numerous and excellent parallel talks by LHC speakers at Hadron2011.
Results from the LHC in the related topics of 
charmonium \cite{gao}  and bottomonium \cite{leonardo}
can be found in two separate plenary talks. However, some important
heavy flavor measurements will be skipped, not because of lack
of interest, but simply because of lack of space. These include:
measurements of many heavy flavor states that have been observed
previously; studies of $CP$-violation and other electro-weak topics;
the search for rare $B$ and $D$ decays;
and all of top-quark physics.

\section{Experimental Issues}

The wealth of results presented here was made possible, in large part, 
by the excellent
performance of the LHC machine. Results discussed in these proceedings
were obtained with up to 40 \unitexp{pb}{-1} data taken in 2010 where
maximum peak luminosities of \scinot{2}{32} \instLunit\ were
observed. Already, at the time of this conference in June 2011, LHC
peak luminosity had been increased by an order of magnitude and more
than 1 \unitexp{fb}{-1} had been delivered to ATLAS and
CMS. Approximately a third of that was seen by LHCb, which limits
instantaneous luminosity to protect its delicate vertex detectors.
As we will see, this rapidly increasing luminosity has important
consequences for the heavy flavor production and spectroscopy programs
at the four experiments, particularly ALICE, ATLAS, and CMS.

Another important factor in obtaining the results that I will discuss
is the excellent performance of the detectors. All of them
recorded collisions from the LHC with more than 90\% efficiency in
2010. The detectors themselves are described in detail in
references \cite{alice-det,atlas-det,cms-det,lhcb-det}. Their
most important attributes for the study of heavy flavors are:
angular acceptance, triggering, tracking, and particle
identification. In order to understand the results produced by the
experiments, it's useful to compare their approaches in each of these
areas. 

An extremely important feature of the experiments
for heavy flavor related measurements is the
range that they cover in $\eta \equiv -\ln [\tan (\theta /2)]$. 
Heavy quarks are produced predominantly in the forward region 
(\abs{\eta}$\gtrsim$1.5)
in proton-proton collisions at LHC energies.
However backgrounds also peak in this area making the extraction of
signals more difficult.
ATLAS and CMS are ``central''
detectors, with muon coverage of \abs{\eta}$<$2.7 and
\abs{\eta}$<$2.4, respectively.
ALICE also has central coverage for electrons, \abs{\eta}$<$0.9, but
detects muons in the forward direction, $-4.0 < \eta < -2.5$.
Finally, LHCb has only forward coverage: $1.9 < \eta < 4.9$.
These acceptances are summarized in Fig.~\ref{fig:accept}.

Of all the elements of the LHC detectors, trigger systems are
perhaps the most critical in constraining the heavy flavor capabilities
of the experiments. 
To deal with the challenge of selecting $\sim$200 Hz of interesting
events from the 40 MHz rate of bunch crossings,
the experiments have constructed
complicated, multi-level trigger systems.
Even with state-of-the-art triggers, though, heavy flavors
present many difficulties.
Heavy flavor events tend to have lower \pT\ scales
than most electro-weak or new physics processes. Thus, their properties
overlap much more strongly with the overwhelming background of light
quark production than do, for example, those of $W$, $Z$, or Higgs
events. This problem is exacerbated as luminosities go up and the
average number of proton-proton collisions per bunch crossing
increases. 

The experiments deal with these problems in a variety of ways.
For those analyses that need inclusive event selections,
``minimum-bias'' or low \pT\ jet triggers must be used, but these gave
acceptable rates only during very early LHC running when luminosities
were orders of magnitude lower than present values. Hence, inclusive
analyses tend to be constrained to only a tiny fraction of the total
LHC running period.
Since heavy hadrons decay semi-leptonically with branching ratios of
\order{10\%}, triggers that are sensitive to leptons (particularly
muons) can be very effective at selecting events containing $B$- and
$D$-hadrons -- but only those decaying to muons.
Additionally, trigger rates for the relatively low \pT\ muons ($<$10
GeV) that tend to be produced in $B$ and $D$ decays quickly become
untenable as luminosities increase. So the most effective single-muon
triggers for heavy flavor physics were eliminated fairly early in the
2010 run. Only low \pT\ dimuon triggers remain active, and even those
are seeing their thresholds gradually creep up.
Finally, LHCb takes advantage of its unique capabilities to construct
displaced vertex triggers at its lowest trigger level (the other
experiments employ displaced vertex triggers at higher levels). 
These triggers
remain unprescaled to the highest luminosity allowed by LHCb and give
it access to fully hadronic decays of heavy hadrons.

Once events are accepted by the trigger system, offline reconstruction
of their properties becomes the most important consideration. Tracking
and particle identification are particularly critical for heavy
hadrons. In the area of tracking two issues often arise in heavy
flavor analyses: 
mass reconstruction and secondary vertex finding.
The accuracy of the first of these depends primarily on precise
measurements of track momenta,
while the second is driven by the spatial (hit) resolution achievable
by individual tracking elements and by the position of the inner-most
tracking element. The vertexing performance of all four detectors is
fairly similar. As an example, each achieves an impact parameter
resolution of $\sim$30 \um\ for tracks in the 5--10 GeV range. Larger
differences are seen in mass resolution, as illustrated in
Fig.~\ref{fig:mass-res}, with LHCb having the clear advantage here.

Finally, particle identification is an important part of most heavy
flavor analyses. All four of the experiments have excellent
capabilities to identify leptons, and each  has some handles on
$\pi/K/p$ separation. However, LHCb makes the most extensive use of hadron
identification in the results presented here. Their RICH detectors
allow them to tag $\sim$95\% of charged kaons with a pion
contamination of only 7\%.

\begin{figure}[htb]
\begin{minipage}[t]{0.495\textwidth}
  \begin{center}
    \includegraphics[width=\textwidth]{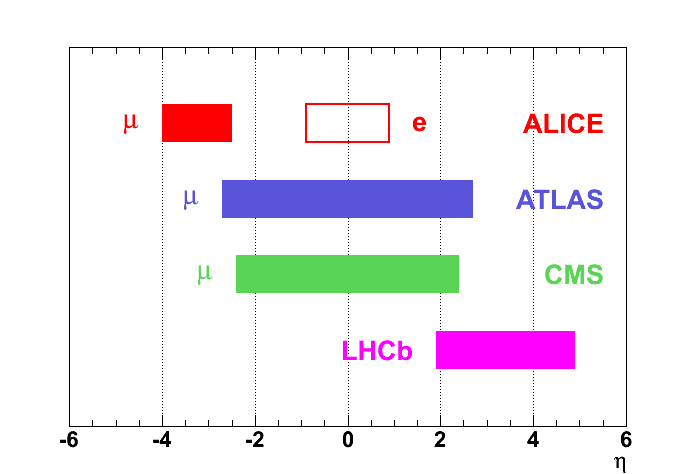}
    \caption{Angular acceptances of the four large LHC experiments.}
    \label{fig:accept}
  \end{center}
\end{minipage}
\hfill
\begin{minipage}[t]{0.495\textwidth}
  \begin{center}
    \includegraphics[width=\textwidth]{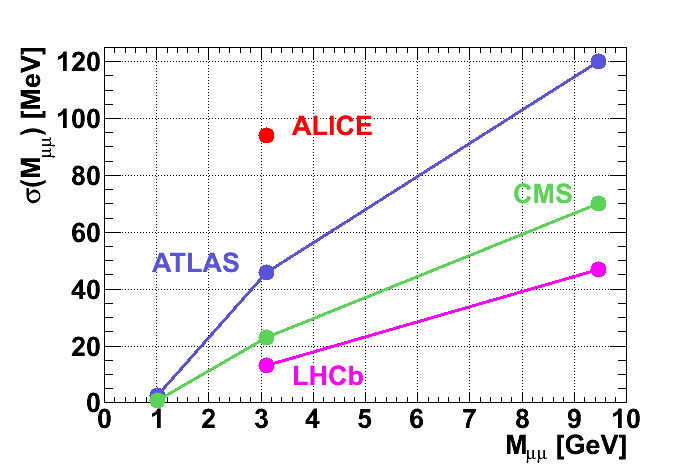}
    \caption{Dimuon mass resolution vs invariant mass for the four
      large LHC experiments.}
    \label{fig:mass-res}
  \end{center}
\end{minipage}
\end{figure}

\section{Heavy Flavor Production}
The study of beauty and charm production at colliders has a long and
interesting history. Because of the relatively high $b$ and $c$ quark
masses, perturbative calculations of their production were expected to
converge rather quickly. However, until the early 2000's, measurements
of $b$ production at the Tevatron and HERA were generally factors of
2--3 higher than Next to Leading Order (NLO) predictions.
The shape 
of the $b$-production spectrum and charm production were much better
described. 
For a more detailed review of the situation see reference
\cite{cacciari}. 
This problem was largely resolved
(see reference \cite{mangano} for a review) by a combination of
experimental and theoretical improvements. 
On the theoretical side, problems due to large $\ln(p_T/m_b)$ terms at
high $b$-quark \pT\ were ameliorated by the use of Fixed Order Next to
Leading Log (FONLL) resummation; and consistent, FONLL, treatment of
fragmentation functions in the calculations was achieved.
Experimentally, collaborations 
made use of these consistent calculations, 
took more care in reporting observations that were less sensitive to
fragmentation/hadronization uncertainties ($B$-hadron and $b$-jet
cross-sections), 
and used updated values of Parton Density Functions and
$\alpha_s$.

The result of these improvements was generally excellent agreement
between measurement and NLO prediction \cite{mangano}.
However, it is important to verify that this agreement is maintained
at the high energies of LHC collisions.
At the present time, our NLO understanding has been incorporated into the
standard, Leading Order (LO) event generators, \Pythia\ \cite{pythia}
and \Herwig\ \cite{herwig} through a variety of interfaces
(\MCatNLO\ \cite{mcatnlo}, \POWHEG\ \cite{powheg}, FONLL \cite{fonll}
are used in the analyses presented here) 
all implementing the calculations in slightly different ways.
Additionally, intermediate implementations of beyond-LO calculations
also exist 
(for example, MadGraph/MadEvent \cite{madgraph} and CASCADE
\cite{cascade})
that allow specific features of those calculations to be probed.
All of these can be used to create detailed predictions that can be
compared directly to experimental data.

\subsection{Inclusive Heavy Flavor Production}
Preliminary, inclusive measurements of heavy flavor production in
\pp\ collisions at \sqrts\ = 7 TeV have
been made by the ALICE and ATLAS collaborations.
Inclusive samples of electrons and muons are selected by both
experiments. ALICE considers electrons in the central region
(\abs{y}$<$0.8) in 2.6 \unitexp{nb}{-1} of data 
and subtracts a mix of ``photonic'' background from 
$\gamma$ conversions, $\pi^0$, and other Dalitz decays based on the
measured $\pi^0$ cross-section. They measure muons in the forward
region ($-$4$< \eta < -$2.5) using a 16.5 \unitexp{pb}{-1} data set. 
These samples are dominated by $b$ and
$c$ decays to leptons. Both measurements agree quite well with FONLL
predictions, as shown (for the electron case) in
Fig.~\ref{fig:HFtoEAlice}.

The ATLAS analysis \cite{atlas-hftolep} uses 1.3--1.4 \unitexp{pb}{-1} of
data selected with single electron and muon triggers of varying
thresholds. Theoretical predictions for Drell-Yan $W/Z/\gamma^*$ are
subtracted resulting in spectra that are dominated by heavy flavor
decays. The collaboration reports differential cross-sections for
electrons in the kinematic range, 
7 $<$ \pT\ $<$ 26 GeV and \abs{\eta} $<$ 2.0; 
and for muons in the range, 
4 $<$ \pT\ $<$ 100 GeV and \abs{\eta} $<$ 2.5.
These measurements are compared to predictions using FONLL, 
\POWHEG +\Pythia , and \POWHEG +\Herwig . Figure \ref{fig:HFtoMuAtlas}
shows the muon result. Good agreement is observed between both
measurements and the FONLL prediction. The generators, 
\POWHEG +\Pythia\ and \POWHEG +\Herwig , also do a reasonably good job
although the \Herwig\ version predicts a significantly lower
cross-section. Interestingly, the ATLAS muon data now has enough reach
in \pT\ to be sensitive to the deviation between the pure NLO and the
FONLL calculations that becomes significant for \pT\ $>$ 35
GeV. The data indicates clearly the need for an NLL resummation at
high \pT .

\begin{figure}[htb]
\begin{minipage}[t]{0.495\textwidth}
  \begin{center}
    \includegraphics[width=\textwidth]{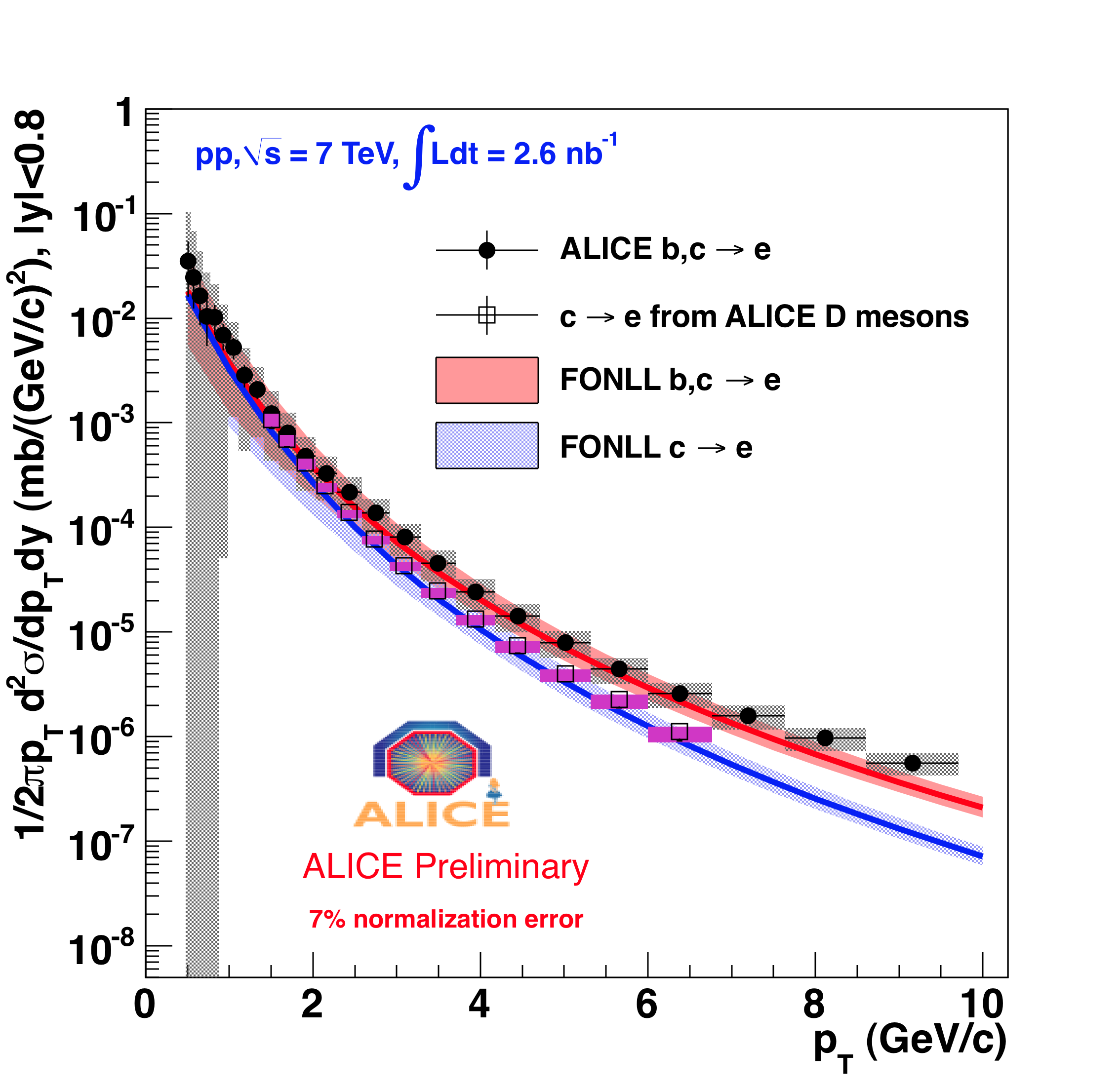}
    \caption{The differential $pp \ra e X$ cross-section at \sqrts\ =
      7 TeV measured by ALICE in \abs{y} $<$ 0.8.}
    \label{fig:HFtoEAlice}
  \end{center}
\end{minipage}
\hfill
\begin{minipage}[t]{0.495\textwidth}
  \begin{center}
    \includegraphics[width=0.75\textwidth]{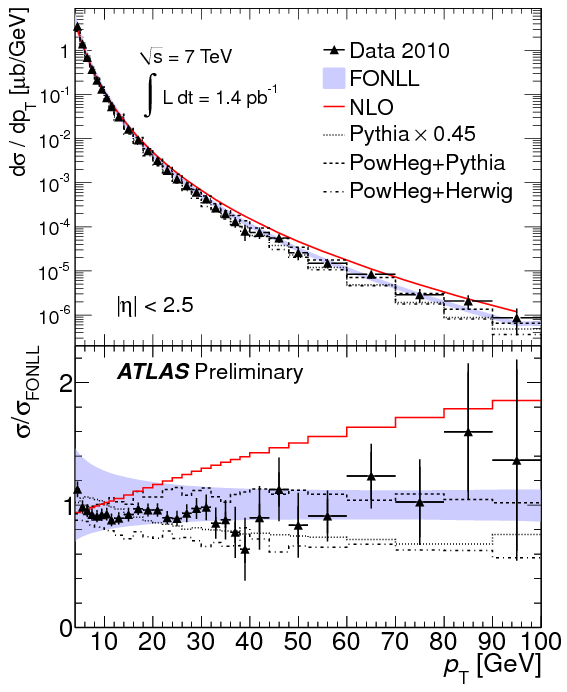}
    \caption{The differential $pp \ra \mu X$ cross-section at
      \sqrts\ = 7 TeV measured by ATLAS in \abs{\eta} $<$ 2.5.}
    \label{fig:HFtoMuAtlas}
  \end{center}
\end{minipage}
\end{figure}

\subsection{Charm Production}
ALICE, ATLAS, and LHCb have all made preliminary measurements of charm
production 
using samples of $D$ mesons collected using micro- or minimum-bias
triggers. The ALICE result is based on samples of 
$D^0 \ra K^- \pi^+$, $D^+ \ra K^- \pi^+ \pi^+$, and 
$D^{*+} \ra D^0 (K^- \pi^+ ) \pi^+$ (and charge conjugate) decays
selected from 1.6 \unitexp{nb}{-1}, about 20\% of the 2010 data. They
measure differential cross-sections in the range 
2 $<$ \pT $<$ 12 GeV in different rapidity ranges that are then
adjusted to \abs{y} $<$ 0.5.
ATLAS uses
$D^+ \ra K^- \pi^+ \pi^+$, $D^{*+} \ra D^0 (K^- \pi^+ ) \pi^+$, and
$D_s^+ \ra \phi (K^+ K^- )$ (and charge conjugate) decays in 1.1
\unitexp{nb}{-1} of data taken during early 2010 running to measure
differential cross-sections in the kinematic range, 
\pT\ $>$ 3.5 GeV (extending to $\sim$40 GeV) and \abs{\eta} $<$ 2.1
\cite{atlas-Dxs}. 
These data contain contributions from both $c$- and $b$-quark
production. However, charm production dominates by approximately a
factor of 20.
Finally, LHCb reconstructs all of the above decay modes using 1.18
\unitexp{nb}{-1} of early data taken with a micro-bias trigger
in the kinematic region, 0 $<$ \pT\ $<$ 8 Gev.
They present differential cross-sections in several rapidity region in 
2 $<$ $y$ $<$ 4.5 \cite{lhcb-charm}.
Charm and beauty components to this data sample are separated using a
fit to the $D$ meson impact parameter distribution.

All differential cross-section results are good agreement with NLO
predictions in a variety of different implementations.
However, uncertainties on these predictions are quite large
and measurements are now limited by systematics.

\subsection{Beauty Production}
Beauty production at the LHC is measured using a wide variety of
different methods. 
ATLAS \cite{atlas-ptrel} and CMS \cite{cms-ptrel} 
separate $b$-production from lighter quark events using the momenta of
muons transverse to a nearby jet's direction ($p_T^{\mathrm{rel}}$) as
a discriminant that is sensitive to the underlying parton's mass.
In samples of 4.8 \unitexp{pb}{-1} collected with low
\pT\ muon+jet triggers (ATLAS),
and 85 \unitexp{nb}{-1} of low \pT\ single-muon triggers (CMS) the
collaborations measure differential cross-sections as a function of
$b$-jet \pT\ (ATLAS) and muon \pT\ (CMS).
Another inclusive method employed by 
ATLAS \cite{atlas-bjet} and CMS \cite{cms-bjet} 
selects $b$-quark events using jets
containing reconstructed secondary vertices
in 3.0 \unitexp{pb}{-1} and 60 \unitexp{nb}{-1}, respectively,
of data taken using a
mixture of minimum-bias and jet triggers.
This method allows
sensitivity to higher values of jet \pT\ (up to 260 GeV) than the
$p_T^{\mathrm{rel}}$ technique.
The substantial $B$-hadron lifetime also provides a handle on
$B$-hadron production in several partially inclusive results. 
LHCb uses the $D^0$ impact parameter distribution of
$pp \ra \mu D^0 X$ candidates in 2.9 \unitexp{nb}{-1} of micro-bias,
and 12.1 \unitexp{nb}{-1} of single-muon trigger data to measure the
$pp \ra H_b X$ cross-section in several bins between 2 $<$ $\eta$ $<$
6 \cite{lhcb-mud0}.
ATLAS \cite{atlas-jpsi}, CMS \cite{cms-jpsi}, and LHCb
\cite{lhcb-jpsi} all use a pseudo-proper time variable in their
\Jpsi\ analyses to measure the $b \ra J/\psi X$ differential
cross-section as well.
Finally, CMS has made $B$-hadron differential cross-section
measurements using the exclusive decay modes
$B^+ \ra \Jpsi (\mumu ) K^+$ \cite{cms-jpsik},
$B_d \ra \Jpsi (\mumu ) K_S^0$ \cite{cms-b0d}, and
$B_s \ra \Jpsi (\mumu ) \phi (K^+ K^-)$ \cite{cms-b0s} (and conjugates)
collected using their dimuon triggers.
These measurements all agree quite well with a
variety of NLO calculations, with the possible exception the exclusive
mode cross-sections where the NLO predictions are consistently lower
than the measurements. But again the uncertainties on the
experimental results are generally dominated by systematic effects
even with the small amounts of integrated luminosity used.

To probe the calculations more deeply, correlations between the
produced $b$ and $\bar{b}$ need to be used. First results in this area
are now available from ATLAS and CMS. 
ATLAS uses its secondary vertexing analysis \cite{atlas-bjet} to
select events with two identified $b$-jets. The dijet invariant mass
that they observe is shown in Fig.~\ref{fig:MbbAtlas}. Agreement with
\POWHEG +\Pythia\ predictions is good across a wide range of dijet
masses. CMS, on the other hand, probes correlations between the $b$
and $\bar{b}$ using angular variables. Their analysis
\cite{cms-bbcor}, which uses 3.1 
\unitexp{nb}{-1} of single-jet trigger data,
searches for events with two reconstructed secondary vertices from
which they reconstruct the opening angle between the two $B$-hadrons
responsible for the vertices. This technique allows access to the events
with very small \qqb{B} opening angles that are particularly sensitive
to NLO effects. An example of their results is shown in
Fig.~\ref{fig:dRbbCms}, where the angular separation,
\dR = $\sqrt{\Delta \phi^2 + \Delta \eta^2}$, between the \qqb{B} pair
observed in CMS data and predicted by various beyond LO calculations 
is plotted compared to the LO \Pythia\ prediction.
None of the higher order predictions do a good job describing the data
in the low \dR\ region where higher order effects are expected to
dominate. 

\begin{figure}[htb]
\begin{minipage}[t]{0.495\textwidth}
  \begin{center}
    \includegraphics[width=\textwidth]{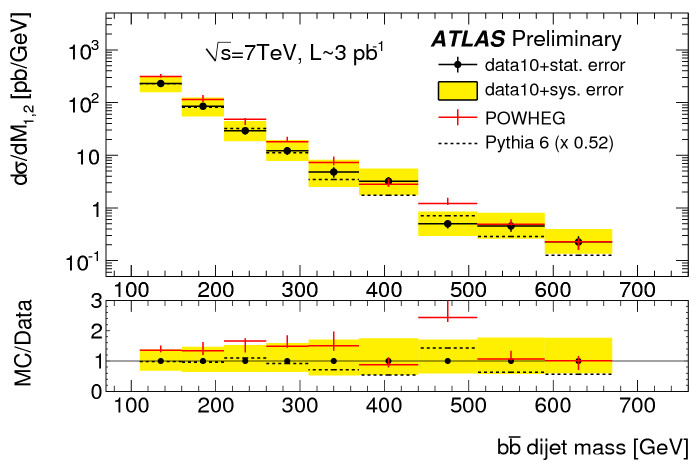}
    \caption{The differential $pp \ra \qqb{b}$ cross-section as a function of
      \qqb{b} dijet invariant mass measured for $b$-jets with \pT
      $>$40 GeV and \abs{y}$<$2.1 at ATLAS.}
    \label{fig:MbbAtlas}
  \end{center}
\end{minipage}
\hfill
\begin{minipage}[t]{0.495\textwidth}
  \begin{center}
    \includegraphics[width=\textwidth]{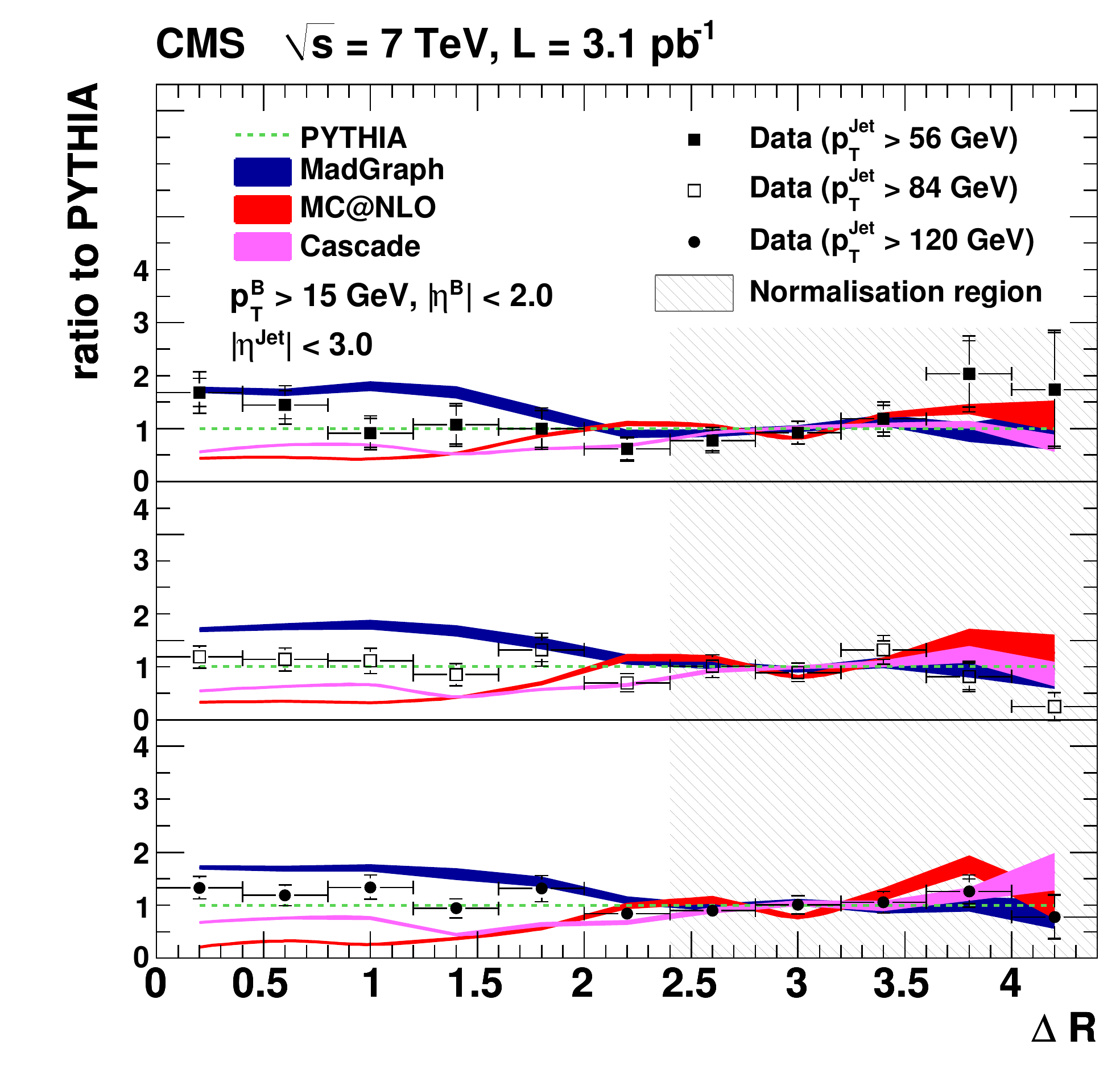}
    \caption{Ratios of differential \qqb{B} cross-sections as a
      function of \dR\ measured by CMS compared to various
      predictions.} 
    \label{fig:dRbbCms}
  \end{center}
\end{minipage}
\end{figure}

\subsection{Heavy Flavor Fragmentation}
Another important component in our understanding of heavy flavor
production is the process of fragmentation of $b$- and $c$-quarks into
beauty and charm hadrons of various flavors. The ATLAS and LHCb
collaborations have produced new measurements in these areas.
The ATLAS study of $D^{(*)}$ meson production \cite{atlas-Dxs} allows
the measurement of several charm fragmentation related parameters
after subtracting predicted $B$-hadron decay contributions from their
measured $D^{(*)}$ meson cross-sections and extrapolating these
\qqb{c} measurements to the full kinematic phase space.
ATLAS derives a strangeness suppression factor, $\gamma_{s/d}$,
which corresponds to the ratio of the total $D_s^{\pm}$ production
cross section from \qqb{c}, $\sigma_{\qqb{c}}^{tot}(D_s^{\pm})$, 
to the sum of $\sigma_{\qqb{c}}^{tot}(D^{*\pm})$ and the part of
$\sigma_{\qqb{c}}^{tot}(D^{\pm})$ that does not arise from $D^{*\pm}$
decays. ATLAS also measures the fraction of $D$ mesons produced in the
vector state, $P_V$, as the ratio of $D^{*\pm}$ to the sum of
$D^{*\pm}$ and $D^{\pm}$ production. Both measurements are in good
agreement with averages of LEP results \cite{atlas-Dxs,lep-cfrac} as
shown in Fig.~\ref{fig:fragm}.

LHCb fragmentation measurements concentrate on the $b$-quark sector.
They have determined the ratio of $b$-quarks fragmenting to hadrons
containing $s$- and $d$-quarks, $f_s/f_d$, using
$B^0 \ra D^- \pi^+$, $B^0 \ra D^- K^+$ and $B_s \ra D_s^- \pi^+$
decays in 35 \unitexp{nb}{-1} of data \cite{lhcb-fsfd}. 
They have also determined the strange quark fraction, 
$f_s / (f_u + f_d )$ and the $\Lambda_b$ baryon fraction,
$f_{\Lambda_b} / (f_u + f_d )$ in $b$-quark fragmentation using
3 \unitexp{pb}{-1}
samples of semi-muonic decays of $B$-hadrons \cite{lhcb-semilep}.
They observe no dependence of the strange quark fraction on \pT\ or
$\eta$, but do see evidence of a linear dependence of the $\Lambda_b$
fraction on \pT\ in all $\eta$ regions they consider.
As can be seen in Fig.~\ref{fig:fragm}, their
results are consistent with \footnote{In
  the case of the $\Lambda_b$ fraction, all results have been adjusted
  to an average \pT\ of 14 GeV.},
but more precise than, the current world
averages \cite{hfag-2010} and individual results from CDF
\cite{cdf-fragm} and LEP \cite{hfag-2010}.

Finally, LHCb has observed a clear signal of 43 $\pm$ 13 events,
in 32.5 \unitexp{pb}{-1} of data,
for
the decay $B_c^{\pm} \ra J/\psi \pi^{\pm}$ \cite{lhcb-bc}. Comparing
this to the production of $B^{\pm} \ra J/\psi K^{\pm}$ yields a ratio,
$\sigma(B_c^+ ) \times B(B_c^+ \ra J/\psi \pi^+ ) \thinspace / \thinspace
\sigma(B^+ ) \times B(B^+ \ra J/\psi K^+ )$ = 
(2.2 $\pm$ 0.8 $\pm$ 0.2)\%,
in good agreement with a prediction made using the BcVegPy generator
\cite{BcVegPy},
(1.4 $\pm$ 0.4 $\pm$ 0.1)\%.

\begin{figure}[htb]
  \begin{center}
    \includegraphics[width=\textwidth]{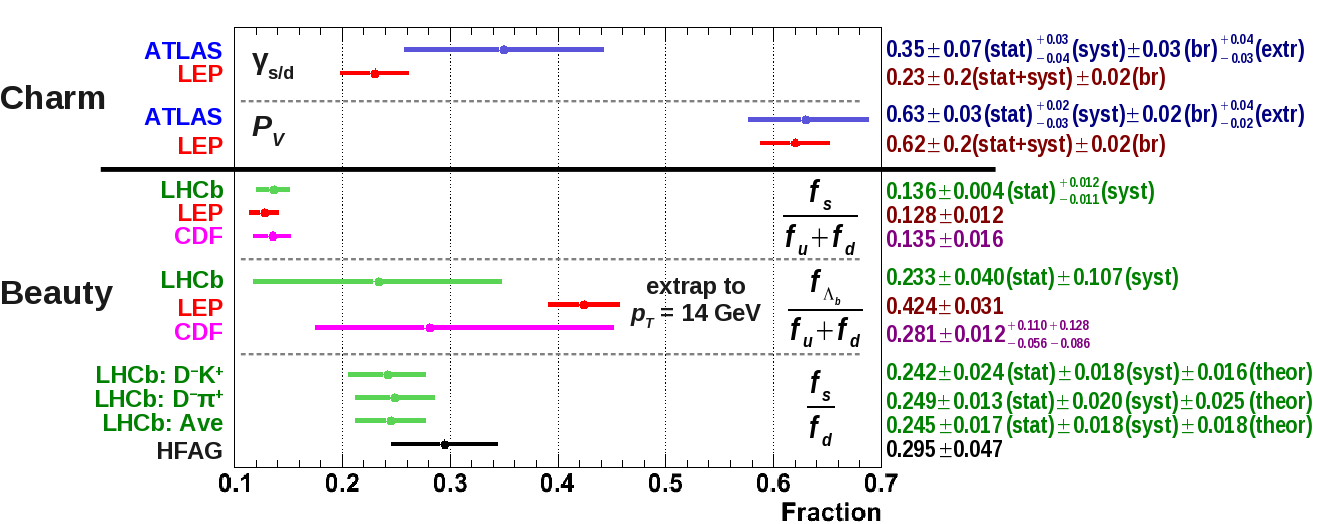}
    \caption{A summary of beauty and charm hadron fragmentation
      fraction measurements. 
      Note: to allow a consistent comparison, the
      $f_{\Lambda_b}$ results are extrapolated to \pT\ = 14 GeV using
      LHCb's parametrization \cite{lhcb-semilep}.
    }
    \label{fig:fragm}
  \end{center}
\end{figure}

\section{Spectroscopy and Exclusive Final States}
All heavy flavor hadrons are produced copiously at the LHC and each of
the large experiments have reconstructed sizable samples of most of
the low-lying states. However, exclusive reconstruction of beauty and
charm hadrons is an area where LHCb is taking a leading role. Their
ability to trigger on hadronic decays and their efficient $\pi /K/p$
identification capabilities, as well as other detector optimizations,
make exclusive final state reconstruction one of their strengths. 

LHCb takes advantage of this strength is several ways. In the areas of
hadron properties, they have been able to identify
several new hadronic decays of \Bs\ mesons with the data they have
taken to
date. They are also using exclusively reconstructed decays as a tool
to study electro-weak symmetry breaking and to search for new physics
through measurements of parameters of the CKM matrix. Of particular
interest here are measurements of the angles 
$\gamma \equiv arg(- V_{ub}^* V_{ud} / v_{cb}^* V_{cd})$
and
$\beta_s \equiv arg(- V_{tb}^* V_{ts} / v_{cb}^* V_{cs})$,
which are currently the most poorly measured of the ``unitarity
triangle'' quantities.

LHCb has observed several new decay modes of the \Bs\ meson. They have
studied \Bs\ decays to excited $c\bar{s}$ states:
$B_s \ra \mu D_{s1} X$ and $B_s \ra \mu D_{s2}^{*} X$ with
$D_{s1}^{+}, D_{s2}^{*+} \ra D^0 (K^- \pi^+) K^+$ \cite{lhcb-dsst}. 
Measurements of the relative fractions of these states allow sensitive
tests of QCD models.
Using data samples of 20 and 3 \unitexp{pb}{-1}, LHCb finds 
\BRat{$\overline{B_s^0} \ra D_{s2}^{*+} X \mu^- \overline{\nu}$} /
\BRat{$\overline{B_s^0} \ra X \mu^- \overline{\nu}$} =
(3.3 $\pm$ 1.0 $\pm$ 0.4)\% and
\BRat{$\overline{B_s^0} \ra D_{s1}^{+} X \mu^- \overline{\nu}$} /
\BRat{$\overline{B_s^0} \ra X \mu^- \overline{\nu}$} =
(5.4 $\pm$ 1.2 $\pm$ 0.4)\%.
The $D_{s1}$ result is in good agreement with a previous measurement
by the D0 collaboration \cite{d0-dsst}, but this is the first
observation of the $D_{s2}^{*}$ decay mode.
These measurements are in general agreement with predictions of 3.2\%
and 5.7\% respectively 
from the ISGW2 model \cite{isgw2}
but slightly higher than quark model expectations of 1.8\% and 2\%
\cite{quarkmodel}.

LHCb has also made a first observation of the decay mode
$B_s^0 \ra K^{*0} \overline{K^{*0}}$,
which proceeds solely through loop $b \ra s$ diagrams in the Standard
Model,
and can be used in the extraction of $\gamma$ and $\beta_s$.
Using 35 \unitexp{fb}{-1} of data they observe a signal with
7$\sigma$ significance, 
as shown in Fig.~\ref{fig:BsKstKst},
and measure
\BRat{$B_s^0 \ra K^{*0} \overline{K^{*0}}$} = 
[1.95 $\pm$ 0.47(stat) $\pm$ 0.51(syst) $\pm$
  0.29($f_d/f_s$)]$\times$10$^{-5}$ \cite{lhcb-kstkst} in reasonable
agreement with the prediction of
(0.79$^{+0.43}_{-0.39}$)$\times$10$^{-5}$ based on QCD factorization
\cite{qcdfact}. 

Measurement of the angle $\gamma$ to date have relied primarily on
$B^- \ra D^{(*)} K^{(*)-}$ decays. 
However, many other modes also
have the potential to contribute. 
These include
$B^0 \ra D^0 K^{*0}$,
$B^- \ra D^0 K^- \pi^+ \pi^-$,
$\overline{B^0} \ra D^+ \pi^- \pi^+ \pi^-$, and
$B_s^0 \ra D_s^+ K^- \pi^+ \pi^-$.
LHCb has taken first steps toward
widening the scope of $\gamma$ measurements by observing several of
these decays or closely related final states.
They measure
\begin{center}
\begin{tabular}{cccc}
  {\bf Mode} & {\bf Events} & {\bf Branching Ratio ($\times 10^4$)} &
  {\bf Ref.}\\
  $\overline{B_s^0} \ra D^0 K^{(*0)}$ & 34.5 $\pm$ 6.9 
    & 4.44 $\pm$ 1.00(stat) $\pm$ 0.55(syst) & \cite{lhcb-bsdk}\\
  & & $\pm$ 0.56($f_s/f_d$) $\pm$ 0.69($B \ra D \rho$) & \\
  $\overline{B^0} \ra D^+ \pi^- \pi^+ \pi^-$ & 1151 $\pm$ 45
    & 61.6 $\pm$ 2.6 $\pm$ 6.9 & \cite{lhcb-xb3pi}\\
  $B^- \ra D^0 \pi^- \pi^+ \pi^-$ & 973 $\pm$ 45
    & 59.6 $\pm$ 2.9 $\pm$ 6.1 & \cite{lhcb-xb3pi}\\
  $\overline{B_s^0} \ra D_s^+ \pi^- \pi^+ \pi^-$ & 139 $\pm$ 24
    & 62.8 $\pm$ 11.0 $\pm$ 12.1 & \cite{lhcb-xb3pi}\\
  $\Lambda_b^0 \ra \Lambda_c^+ \pi^- \pi^+ \pi^-$ & 165 $\pm$ 18
    & 122 $\pm$ 14 $\pm$ 46 & \cite{lhcb-xb3pi}\\
\end{tabular}
\end{center}
which represent either first observations of these final states (the
first and last measurements listed above) or significant improvements
over current world averages.

Finally, LHCb was the first to observe the decay
$B_s \ra J/\psi [\mumu ] f_0 (980) [\pi^+ \pi^-]$ \cite{lhcb-psif0}.
As shown in Fig.~\ref{fig:BsPsiF0} they see a clear peak in the
$J/\psi \pi^+ \pi^-$ distribution at the \Bs\ mass
in 33 \unitexp{pb}{-1} of data.
They measure
$\Gamma[ B_s^0 \ra J/\psi f_0 (\pi^+ \pi^-) ] / 
\Gamma[ B_s^0 \ra J/\psi \phi (K^+ K^-) ]$ =
0.252$^{+0.046 \thinspace + 0.027}_{-0.032 \thinspace -0.033}$ 
in good agreement with the later D0 measurement of
0.210 $\pm$ 0.032 $\pm$ 0.036 \cite{d0-psif0}.
This decay is similar to $B_s \ra J/\psi \phi$, which is commonly used
to determine the angle $\beta_s$.
However, since the $J/\psi f_0$ mode consists of a single $CP$-odd
eigenstate, it can be used to extract $\beta_s$ without having to rely
on the complicated angular analysis necessary to disentangle the $CP$
states in the $J/\psi \phi$ mode.

\begin{figure}[htb]
\begin{minipage}[t]{0.495\textwidth}
  \begin{center}
    \includegraphics[width=\textwidth]{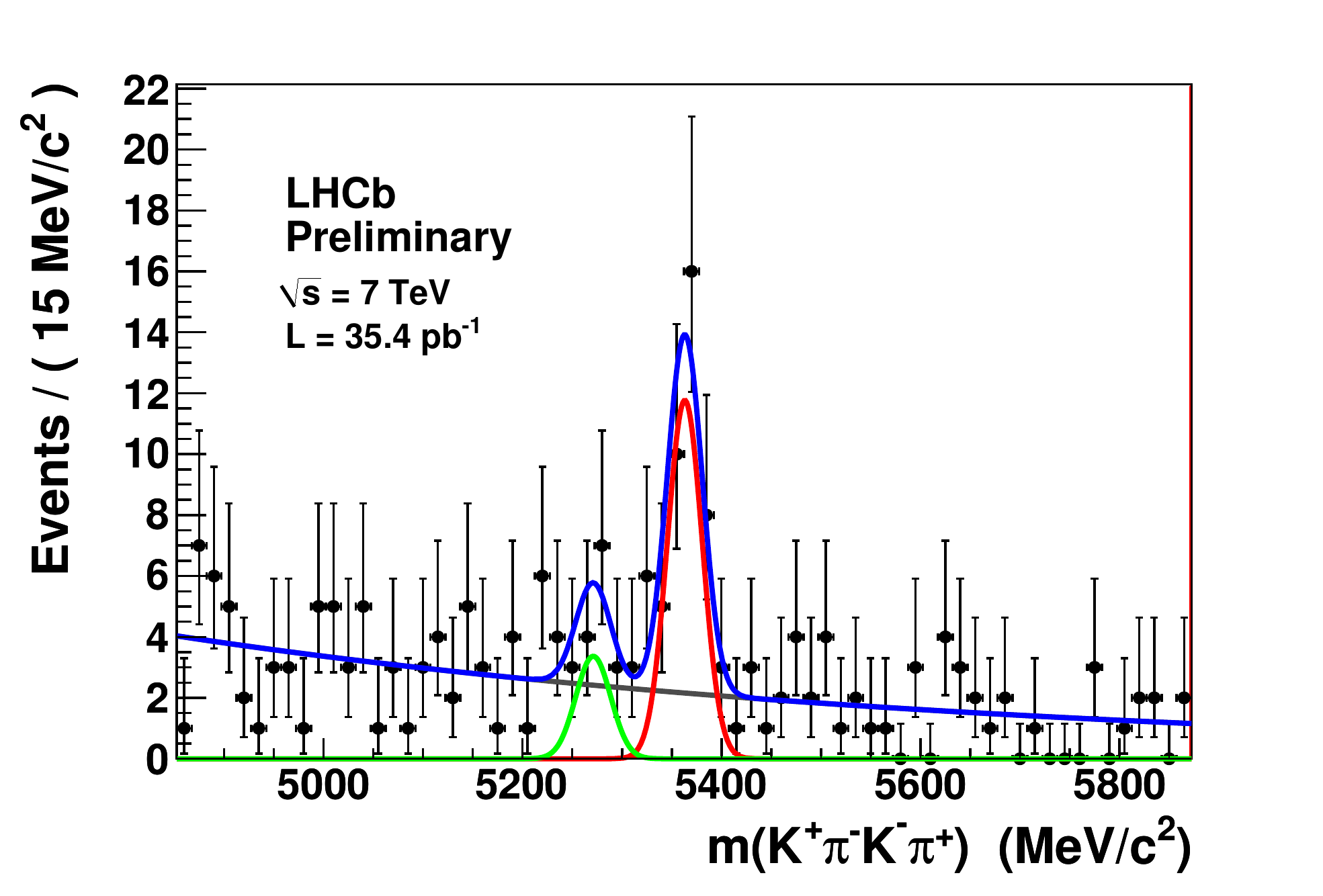}
    \caption{The $K^+ \pi^- K^- \pi^+$ invariant mass distribution
      measured by LHCb  in their $B_s \ra K^{*0} \overline{K^{*0}}$ analysis
      \cite{lhcb-kstkst}}
    \label{fig:BsKstKst}
  \end{center}
\end{minipage}
\hfill
\begin{minipage}[t]{0.495\textwidth}
  \begin{center}
    \includegraphics[width=\textwidth]{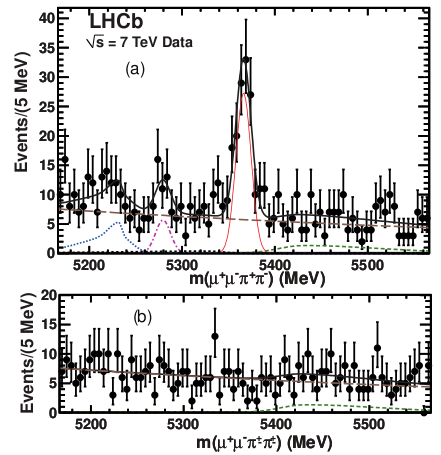}
    \caption{The $J/\psi \pi^+ \pi^-$ invariant mass distribution
      measured by LHCb  in their $B_s \ra J/\psi f_0$ analysis
      \cite{lhcb-psif0} for right sign (top) and wrong sign $\pi \pi$
      combinations, where the $\pi \pi$ mass is required to lie within
      90 MeV of the $f_0(980)$ mass.}
    \label{fig:BsPsiF0}
  \end{center}
\end{minipage}
\end{figure}

\section{Conclusions}
The first year of LHC running has produced a wealth of results in the
heavy flavor sector from ALICE, ATLAS, CMS, and LHCb. Each of the
experiments has made measurements of open beauty and charm production. In
general, the good agreement between data and NLO QCD predictions that
was observed at the Tevatron continues to hold at higher LHC
energies although most measurements are already systematics
limited. Upon considering those measurements that are particularly
sensitive to the details of the calculations, though,  
some areas of disagreement seem to be appearing.
These are most evident in 
exclusive $b$ production and in the angular correlations observed
between $b$ and $\bar{b}$ jets.
Unfortunately (for these measurements), the rapidly increasing LHC
luminosity requires all the experiments to move away from the
inclusive, low \pT\ triggers that have provided the bulk of the data
for production studies. In the future, these studies will need to
shift focus, primarily to rely on those final states that are
accessible using dimuon 
triggers (for example, $\Lambda_b \ra J/\psi \Lambda$). Even for
these, muon \pT\ thresholds will steadily increase. Clearly, the next
round of heavy flavor measurements will have to use
different techniques than those described here.

Trigger considerations also affect the areas of exclusive final states
and spectroscopy for ATLAS and CMS.
However, the LHCb  collaboration is
truly starting to hit its stride here. 
They have already made several first
observations of heavy flavor decay modes and are now beginning to have
enough data to probe $CP$ violation using a wide variety of techniques.

The next few years will be challenging ones for the heavy flavor
efforts of the four large LHC experiments. 
Overcoming the difficulties posed by increasing luminosity and
the demands of ever higher precision will require both perseverance
and cleverness. The excellent results presented here, though, show
that ALICE, ATLAS, CMS, and LHCb are up to the task. Look forward to
even more interesting heavy flavor results from the LHC at the next
meeting of this conference!

\acknowledgements{%
  I am grateful to my colleagues on ALICE, ATLAS, CMS, and LHCb for
  their helpful input to this article. Thanks also to the local
  organizers for an interesting and enjoyable conference.
}


%

}  



\begin{thebibliography}{99}

\bibitem{gao} Y.~Gao, these proceedings.

\bibitem{leonardo} N.~Leonardo, these proceedings.

\bibitem{alice-det}
 K.~Aamodt \etal\ [ALICE Collaboration],
  \Journal{\JINST}{3}{S08002}{2008}.

\bibitem{atlas-det}
  G.~Aad \etal\ [ATLAS Collaboration],
  \Journal{\JINST}{3}{S08003}{2008}.

\bibitem{cms-det}
  R.~Adolphi \etal\ [CMS Collaboration],
  \Journal{\JINST}{3}{S08004}{2008}.

\bibitem{lhcb-det}
  A.~A.~Alves \etal\ [LHCb Collaboration],
  \Journal{\JINST}{3}{S08005}{2008}.

\bibitem{cacciari}
  M.~Cacciari,
  \arXivOld{hep-ph}{0407187}.

\bibitem{mangano}
  M.~Mangano,
  \arXivOld{hep-ph}{0411020}.

\bibitem{pythia}
  T.~Sjostrand, S.~Mrenna, P.~Z.~Skands,
  \Journal{\JHEP}{05}{026}{2006},
  \href{http://arxiv.org/abs/hep-ph/0603175}{\arXivOld{hep-ph}{0603175}}.

\bibitem{herwig}
  G.~Corcella \etal ,
  \Journal{\JHEP}{01}{010}{2001},
  \href{http://arxiv.org/abs/hep-ph/0011363}{\arXivOld{hep-ph}{0011363}}.

\bibitem{mcatnlo}
  S.~Frixione, B.~R.~Webber,
  \Journal{\JHEP}{06}{029}{2002},
  \href{http://arxiv.org/abs/hep-ph/0204244}{\arXivOld{hep-ph}{0204244}}.\\
  S.~Frixione, P.~Nason, B.~R.~Webber,
  \Journal{\JHEP}{08}{007}{2003},
  \href{http://arxiv.org/abs/hep-ph/0305252}{\arXivOld{hep-ph}{0305252}}.

\bibitem{powheg}
 P.~Nason,
  \Journal{\JHEP}{11}{040}{2004},
  \href{http://arxiv.org/abs/hep-ph/0409146}{\arXivOld{hep-ph}{0409146}}.\\
  S.~Frixione, P.~Nason, G.~Ridolfi,
  \Journal{\JHEP}{09}{126}{2007},
  \href{http://arxiv.org/abs/0707.3088}{\arXiv{0707}{3088[hep-ph]}}.\\
  S.~Frixione, P.~Nason, C.~Oleari,
  \Journal{\JHEP}{11}{070}{2007},
  \href{http://arxiv.org/abs/0709.2092}{\arXiv{0709}{2092[hep-ph]}}.\\
 S.~Alioli, P.~Nason, C.~Oleari, E.~Re,
  \Journal{\JHEP}{06}{043}{2010},
  \href{http://arxiv.org/abs/1002.2581}{\arXiv{1002}{2581[hep-ph]}}.

\bibitem{fonll}
  M.~Cacciari, P.~Nason, C.~Oleari,
  \Journal{\JHEP}{04}{006}{2006},
  \href{http://arxiv.org/abs/hep-ph/0510032}{\arXivOld{hep-ph}{0510032}}.

\bibitem{cascade}
  H.~Jung, G.~P.~Salam,
  \Journal{\EPJ }{C19}{351}{2001},
  \href{http://arxiv.org/abs/hep-ph/0012143}{\arXivOld{hep-ph}{0012143}}.

\bibitem{madgraph}
  J.~Alwall, M.~Herquet, F.~Maltoni, O.~Mattelaer, T.~Stelzer,
  \Journal{\JHEP}{06}{128}{2011},
  \href{http://arxiv.org/abs/1106.0522}{\arXiv{1106}{0522[hep-ph]}}.

\bibitem{atlas-hftolep}
  G.~Aad \etal\ [ATLAS Collaboration],
  \href{http://arxiv.org/abs/1109.0525}{\arXiv{1109}{0525[hep-ex]}}.

\bibitem{atlas-Dxs}
  The ATLAS Collaboration,
  \href{http://cdsweb.cern.ch/record/1336746}{ATLAS-CONF-2011-017}.

\bibitem{lhcb-charm}
  The LHCb Collaboration,
  \href{http://cdsweb.cern.ch/record/1311236}{LHCb-CONF-2010-013}.

\bibitem{atlas-ptrel}
  The ATLAS Collaboration,
  \href{http://cdsweb.cern.ch/record/1343733}{ATLAS-CONF-2011-057}

\bibitem{cms-ptrel}
  V.~Khachatryan \etal\ [CMS Collaboration],
  \Journal{\JHEP}{1103}{090}{2011},
  \href{http://arxiv.org/abs/1101.3512}{\arXiv{1101}{3512[hep-ex]}}.

\bibitem{atlas-bjet}
  The ATLAS Collaboration,
  \href{http://cdsweb.cern.ch/record/1342571}{ATLAS-CONF-2011-056}.

\bibitem{cms-bjet}
  The CMS Collaboration,
  \href{http://cdsweb.cern.ch/record/1280454}{CMS-PAS-BPH-10-009}.

\bibitem{lhcb-mud0}
  R.~Aaij \etal\ [LHCb Collaboration],
  \Journal{\PhysLett}{B694}{209}{2010},
  \href{http://arxiv.org/abs/1009.2731}{\arXiv{1009}{2731[hep-ex]}}.

\bibitem{atlas-jpsi}
  G.~Aad \etal\ [ATLAS Collaboration],
  \Journal{\NuclPhys}{B850}{387}{2011},
  \href{http://arxiv.org/abs/1104.3038}{\arXiv{1104}{3038[hep-ex]}}.

\bibitem{cms-jpsi}
  V.~Khachatryan \etal\ [CMS Collaboration],
  \Journal{\EPJ}{C71}{1575}{2011},
  \href{http://arxiv.org/abs/1011.4193}{\arXiv{1011}{4193[hep-ex]}}.

\bibitem{lhcb-jpsi}
  R.~Aaij \etal\ [LHCb Collaboration],
  \Journal{\EPJ}{C71}{1645}{2011},
  \href{http://arxiv.org/abs/1103.0423}{\arXiv{1103}{0423[hep-ex]}}.

\bibitem{cms-jpsik}
  V.~Khachatryan \etal\ [CMS Collaboration],
  \Journal{\PRL}{106}{112001}{2011},
  \href{http://arxiv.org/abs/1101.0131}{\arXiv{1101}{0131[hep-ex]}}.

\bibitem{cms-b0d}
  S.~Chatrchyan \etal\ [CMS Collaboration],
  \Journal{\PRL}{106}{252001}{2011},
  \href{http://arxiv.org/abs/1104.2892}{\arXiv{1104}{2892[hep-ex]}}.

\bibitem{cms-b0s}
  S.~Chatrchyan \etal\ [CMS Collaboration],
  \Journal{\PhysRev}{D84}{052008}{2011},
  \href{http://arxiv.org/abs/1106.4048}{\arXiv{1106}{4048[hep-ex]}}.

\bibitem{cms-bbcor}
  V.~Khachatryan \etal\ [CMS Collaboration],
  \Journal{\JHEP}{1103}{136}{2011},
  \href{http://arxiv.org/abs/1102.3194}{\arXiv{1102}{3194[hep-ex]}}.

\bibitem{lep-cfrac}
  L.~Gladilin,
  \href{http://arxiv.org/abs/hep-ex/9912064}{\arXivOld{hep-ex}{9912064}}.

\bibitem{lhcb-fsfd}
  The LHCb Collaboration,
  \href{http://cdsweb.cern.ch/record/1333395}{LHCb-CONF-2011-013}.

\bibitem{lhcb-semilep}
  The LHCb Collaboration,
  \href{http://cdsweb.cern.ch/record/1356181}{LHCb-CONF-2011-028}.

\bibitem{hfag-2010}
  D.~Asner \etal\ [Heavy Flavor Averaging Group],
  \href{http://arxiv.org/abs/1010.1589}{\arXiv{1010}{1589[hep-ex]}}.

\bibitem{cdf-fragm}
  T.~Aaltonen \etal\ [CDF Collaboration],
  \Journal{\PhysRev}{D77}{072003}{2008},
  \href{http://arxiv.org/abs/0801.4375}{\arXiv{0801}{4375[hep-ex]}}.

\bibitem{lhcb-bc}
  The LHCb Collaboration,
  \href{http://cdsweb.cern.ch/record/1336254}{LHCb-CONF-2011-017}.

\bibitem{BcVegPy}
  C.~-H.~Chang, C.~Driouichi, P.~Eerola, X.~G.~Wu,
  \Journal{\CompPhysComm}{159}{192}{2004},
  \href{http://arxiv.org/abs/hep-ph/0309120}{\arXivOld{hep-ph}{0309120}}.
  C.~-H.~Chang, J.~-X.~Wang, X.~-G.~Wu,
  \Journal{\CompPhysComm}{175}{624}{2006},
  \href{http://arxiv.org/abs/hep-ph/0604238}{\arXivOld{hep-ph}{0604238}}.

\bibitem{lhcb-dsst}
  R.~Aaij \etal\ [LHCb Collaboration],
  \Journal{\PhysLett}{B698}{14}{2011},
  \href{http://arxiv.org/abs/1102.0348}{\arXiv{1102}{0348[hep-ex]}}.

\bibitem{d0-dsst}
  V.~M.~Abazov \etal\ [D0 Collaboration],
  \Journal{\PRL}{102}{051801}{2009},
  \href{http://arxiv.org/abs/0712.3789}{\arXiv{0712}{3789[hep-ex]}}.

\bibitem{isgw2}
  D.~Scora, N.~Isgur,
  \Journal{\PhysRev}{D52}{2783}{1995},
  \href{http://arxiv.org/abs/hep-ph/9503486}{\arXivOld{hep-ph}{9503486}}.

\bibitem{quarkmodel}
  H.~B.~Mayorga, A.~Moreno Briceno, J.~H.~Munoz,
  \Journal{J.\ Phys.\ }{G29}{2059}{2003},
  \href{http://arxiv.org/abs/hep-ph/0209032}{\arXivOld{hep-ph}{0209032}}.

\bibitem{lhcb-kstkst}
  The LHCb Collaboration,
  \href{http://cdsweb.cern.ch/record/1344131}{LHCb-CONF-2011-019}.

\bibitem{qcdfact}
  M.~Beneke, J.~Rohrer, D.~Yang,
  \Journal{\NuclPhys} {B774}{64}{2007},
  \href{http://arxiv.org/abs/hep-ph/0612290}{\arXivOld{hep-ph}{0612290}}.

\bibitem{lhcb-bsdk}
  The LHCb Collaboration,
  \href{http://cdsweb.cern.ch/record/1329858}{LHCb-CONF-2011-008}.

\bibitem{lhcb-xb3pi}
  The LHCb Collaboration,
  \href{http://cdsweb.cern.ch/record/1329855}{LHCb-CONF-2011-007}.

\bibitem{lhcb-psif0}
  R.~Aaij \etal\ [LHCb Collaboration],
  \Journal{\PhysLett}{B698}{115}{2011},
  \href{http://arxiv.org/abs/1102.0206}{\arXiv{1102}{0206[hep-ex]}}.

\bibitem{d0-psif0}
  The D0 Collaboration,
  \href{http://www-d0.fnal.gov/Run2Physics/WWW/results/prelim/B/B62/}%
       {D0 Note 6152} (2011).

\end{thebibliography}
\end{document}